\documentclass[twocolumn,secnumarabic,amssymb, nobibnotes, aps, prd, superscriptaddress]{revtex4-2}
\usepackage{algorithm}
\usepackage[noend]{algpseudocode}

\usepackage{scalerel}
\usepackage{tikz}
\usetikzlibrary{svg.path}
\definecolor{orcidlogocol}{HTML}{A6CE39}
\tikzset{
  orcidlogo/.pic={
    \fill[orcidlogocol] svg{M256,128c0,70.7-57.3,128-128,128C57.3,256,0,198.7,0,128C0,57.3,57.3,0,128,0C198.7,0,256,57.3,256,128z};
    \fill[white] svg{M86.3,186.2H70.9V79.1h15.4v48.4V186.2z}
                 svg{M108.9,79.1h41.6c39.6,0,57,28.3,57,53.6c0,27.5-21.5,53.6-56.8,53.6h-41.8V79.1z M124.3,172.4h24.5c34.9,0,42.9-26.5,42.9-39.7c0-21.5-13.7-39.7-43.7-39.7h-23.7V172.4z}
                 svg{M88.7,56.8c0,5.5-4.5,10.1-10.1,10.1c-5.6,0-10.1-4.6-10.1-10.1c0-5.6,4.5-10.1,10.1-10.1C84.2,46.7,88.7,51.3,88.7,56.8z};
  }
}

\newcommand\orcidicon[1]{\href{https://orcid.org/#1}{\mbox{\scalerel*{
\begin{tikzpicture}[yscale=-1,transform shape]
\pic{orcidlogo};
\end{tikzpicture}
}{|}}}}

\usepackage{hyperref}
\usepackage{multirow}

\definecolor{orcidlogocol}{HTML}{A6CE39}
\newcommand{\orcid}[1]{\href{https://orcid.org/#1}{\textcolor[HTML]{A6CE39}{\aiOrcid}}}

\begin{document}

\title{Online Bayesian Optimization for a Recoil Mass Separator}


\author{S. A. Miskovich\orcidicon{0000-0002-3302-838X}}
\email{smiskov@slac.stanford.edu}
\altaffiliation{SLAC National Accelerator Laboratory, Menlo Park, 94025 California}
\affiliation{Department of Physics and Astronomy, Michigan State University, East Lansing, MI 48824}
\affiliation{National Superconducting Cyclotron Laboratory, Michigan State University, East Lansing, MI 48824}
\affiliation{The Joint Institute for Nuclear Astrophysics, Michigan State University, East Lansing, MI 48824}

\author{F. Montes}
\affiliation{National Superconducting Cyclotron Laboratory, Michigan State University, East Lansing, MI 48824}
\affiliation{The Joint Institute for Nuclear Astrophysics, Michigan State University, East Lansing, MI 48824}

\author{G. P. A. Berg}
\affiliation{Department of Physics, University of Notre Dame, Notre Dame, IN 46556}
\affiliation{The Joint Institute for Nuclear Astrophysics, University of Notre Dame, Notre Dame, IN 46556}

\author{J. Blackmon}
\affiliation{Department of Physics and Astronomy, Louisiana State University, Baton Rouge, LA 70803}

\author{K. A. Chipps}
\affiliation{Physics Division, Oak Ridge National Laboratory, Oak Ridge, TN 37831}

\author{M. Couder}
\affiliation{Department of Physics, University of Notre Dame, Notre Dame, IN 46556}
\affiliation{The Joint Institute for Nuclear Astrophysics, University of Notre Dame, Notre Dame, IN 46556}

\author{C. M. Deibel}
\affiliation{Department of Physics and Astronomy, Louisiana State University, Baton Rouge, LA 70803}

\author{K. Hermansen\orcidicon{0000-0002-1707-0844}}
\affiliation{Department of Physics and Astronomy, Michigan State University, East Lansing, MI 48824}
\affiliation{National Superconducting Cyclotron Laboratory, Michigan State University, East Lansing, MI 48824}
\affiliation{The Joint Institute for Nuclear Astrophysics, Michigan State University, East Lansing, MI 48824}

\author{A. A. Hood}
\affiliation{Cyclotron Institute, Texas A\&M University, College Station, TX 77843}

\author{R. Jain}
\affiliation{Department of Physics and Astronomy, Michigan State University, East Lansing, MI 48824}
\affiliation{National Superconducting Cyclotron Laboratory, Michigan State University, East Lansing, MI 48824}
\affiliation{The Joint Institute for Nuclear Astrophysics, Michigan State University, East Lansing, MI 48824}

\author{T. Ruland}
\affiliation{Department of Physics and Astronomy, Louisiana State University, Baton Rouge, LA 70803}

\author{H. Schatz}
\affiliation{Department of Physics and Astronomy, Michigan State University, East Lansing, MI 48824}
\affiliation{National Superconducting Cyclotron Laboratory, Michigan State University, East Lansing, MI 48824}
\affiliation{The Joint Institute for Nuclear Astrophysics, Michigan State University, East Lansing, MI 48824}

\author{M. S. Smith}
\affiliation{Physics Division, Oak Ridge National Laboratory, Oak Ridge, TN 37831}

\author{P. Tsintari}
\affiliation{Department of Physics, Central Michigan University, Mt Pleasant, MI 48859}

\author{L. Wagner}
\affiliation{Department of Physics and Astronomy, Michigan State University, East Lansing, MI 48824}
\affiliation{National Superconducting Cyclotron Laboratory, Michigan State University, East Lansing, MI 48824}
\affiliation{The Joint Institute for Nuclear Astrophysics, Michigan State University, East Lansing, MI 48824}

\date{January 2022}

\begin{abstract}
The SEparator for CApture Reactions (SECAR) is a next-generation recoil separator system at the Facility for Rare Isotope Beams (FRIB) designed for the direct measurement of capture reactions on unstable nuclei in inverse kinematics. To maximize the performance of this system, stringent requirements on the beam alignment to the central beam axis and on the ion-optical settings need to be achieved. These can be difficult to attain through manual tuning by human operators without potentially leaving the system in a sub-optimal and irreproducible state. In this work, we present the first development of online Bayesian optimization with a Gaussian process model to tune an ion beam through a nuclear astrophysics recoil separator. We show that this method achieves small incoming angular deviations (\textless 1 mrad) in an efficient and reproducible manner that is at least 3 times faster than standard hand-tuning. Additionally, we present a Bayesian method for experimental optimization of the ion optics, and show that it validates the nominal theoretical ion-optical settings of the device, and improves the mass separation by 32\% for some beams.
\end{abstract}

\maketitle

\section{Introduction} \label{intro}

The SEparator for CApture Reactions (SECAR) is a next-generation recoil separator system installed at the  ReA3 accelerator \cite{Kester2011} at the National Superconducting Cyclotron Laboratory (NSCL) and the Facility for Rare Isotope Beams (F\-RIB) at Michigan State University (MSU). It is optimized for direct measurements in inverse kinematics of low energy capture reaction rates of proton and alpha particles on short-lived proton-rich nuclei that are crucial to addressing open questions in nuclear astrophysics regarding explosive stellar scenarios including nova explosions, X-ray bursts, and supernovae \cite{Berg2018}. SECAR has been undergoing beam commissioning during which beam tuning procedures and optimizations have been developed for the device.  

To achieve precision proton- and alpha-capture measurements, SECAR uses a series of electrostatic and magnetic elements to maximize the transmission of heavy reaction product particles ("recoils") to a set of detectors,  while simultaneously maximizing the rejection of unreacted beam particles. Although these two particle types have different masses, their nearly identical momenta make this rejection challenging. To achieve an ultimate beam rejection goal of 10\textsuperscript{-13} (not including additional rejection using the detector systems), SECAR brings the recoils to a focus in two mass-dispersed planes along the separator and steers the unreacted beam particles into slits. SECAR consists of eight dipole magnets, 15 quad\-ru\-pole magnets, three hex\-a\-pole magnets, one oct\-u\-pole magnet, and two velocity filters. For a detailed description of the SECAR design, see \cite{Berg2018}. 

Separation of unreacted beam and recoil particles requires a carefully tuned beam that is aligned along the separator's ion-optical axis, and finely tuned magnetic fields that maximize mass resolution and unreacted beam rejection. Once the system has been tuned to the electric and magnetic rigidity of the recoil particles, the rejection of unreacted beam particles is maximized by minimizing the beam spot size at the mass selection focal planes and using slits to remove unreacted particles. 

\begin{figure}
    \centering
    \includegraphics[width=8.6cm]{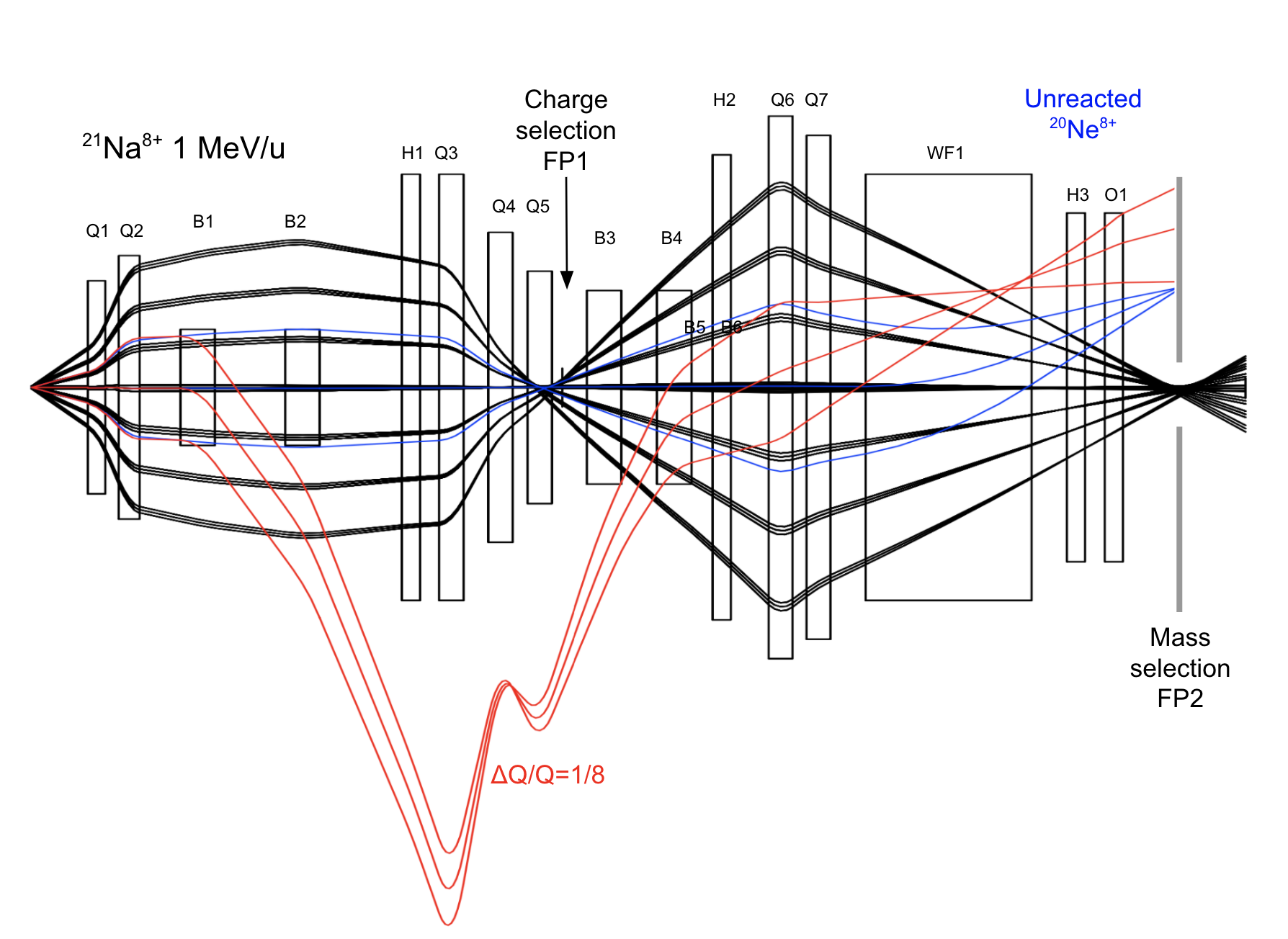}
    \caption{COSY Infinity ion optics calculation (x-plane only) of a proton-capture reaction on \textsuperscript{20}Ne\textsuperscript{8+}. SECAR elements are shown up to the first mass separation slits. The recoils (\textsuperscript{21}Na\textsuperscript{8+}) are shown in black for different outgoing energies and angles, recoils of charge $Q$ = 9+ are shown in red, and unreacted \textsuperscript{20}Ne\textsuperscript{8+} of similar magnetic rigidity is shown in blue.}
    \label{fig:cosy}
\end{figure}

SECAR's ion-optical design is illustrated in Figure \ref{fig:cosy} where 100 rays within the acceptance are traced up to the first mass selection slits using the COSY INFINITY beam simulation code \cite{MAKINO2006346}. Once a single charge state of the incoming beam reaches the SECAR target \cite{CHIPPS2014,SCHMIDT2018}, charge-changing processes cause the production of multiple charge states. Most of the projectiles exit the target unreacted, having a charge state that is dependent on the target material, beam species, and beam energy. Recoils produced via capture reactions in the target also have such interactions as they exit the target. Hence both unreacted beam and recoils have charge state distributions as they enter the separator. A first set of slits selects a single recoil charge state at the first focal plane (FP1) letting through a mix comprised of recoils and unreacted beam particles of nearly identical magnetic rigidity (black and blue rays). The second set of slits stops the unreacted beam of similar momentum (blue rays) at the mass-dispersive focus (FP2).

SECAR's ion-optical design is such that recoils have a small beam spot at the mass selection FP2 while maintaining a large physical separation from the unreacted beam particles. The mass resolution at FP2 is defined as the ratio of the mass dispersion to the magnification multiplied by the full object size (\textit{i.e.}, divided by the beam size at FP2). Therefore to achieve a good beam rejection, the ratio of the distance $\Delta x$ between the recoil focus (black) and the unreacted beam focus (blue) to the beam spot width $w$ is maximized. This allows for the slits at that focus to be closed tightly around the recoils and reduce transmission through the separator from unwanted beam particles (usually referred to as leaky beam). Both $w$ and $\Delta x$ are characteristic of the reaction and magnet settings: $\Delta x$ is related to the fractional mass difference between the beam and the recoil, while $w$ depends on the size of the incident beam, as well as on the quad\-ru\-pole, hex\-a\-pole and oct\-u\-pole settings (in the case shown in Figure \ref{fig:cosy}, quad\-ru\-poles Q1 to Q7, hex\-a\-poles H1 to H3, and oct\-u\-pole O1).

Extensive ion-optical simulations were performed to develop a SECAR design that meets all requirements needed to simultaneously achieve a high transmission of recoils and a high rejection of unreacted projectiles \cite{Berg2018}. For the system to reach the design performance, careful tuning of magnets and velocity filters is required to achieve the stringent conditions imposed on the incoming beam angle, position and size at the SECAR target position (first two rows in Table \ref{table:beam req}). 

The process of manually adjusting the incoming beam properties as well as the nominal SECAR magnet settings to optimize beam rejection in the actual system can take a significant amount of beam time. Tasks such as visual checks of tune quality are operator dependent, introducing subjectivity and bias to the process, and might leave the device in a sub-optimal and irreproducible state. A more robust solution can be achieved with an automated optimizer that enhances reproducibility, ensures objectivity when assessing tune quality, and operates with an efficiency that surpasses the speed of manual tuning when searching for the optimal parameters to achieve SECAR performance targets for each experiment.
\begin{table*}
    \centering
    \begin{tabular}{c c c}
    \hline
     & \textbf{Property} & \textbf{Requirement} \\
    \hline
     \multirow{2}{10em}{\textbf{Incoming Beam Angle and Position}} & Maximum size at target  &	1.5$\times$1.5 mm   \\ [0.5ex]
     & Angular deviation at target &  \textless 1 mrad   \\ [0.5ex]
     \hline\\ [-2ex]
     \multirow{2}{10em}{\textbf{Beam Rejection}} & Width of mass separation slits & Minimized \\ [0.5ex]
     & Transmission to final focus within energy acceptance & 100\%   \\ [0.5ex]
    \end{tabular}
         \caption{Beam requirements for optimal recoil separation in SECAR.}
    \label{table:beam req}
\end{table*}

Machine learning model-dependent optimization methods have been successfully applied at other facilities to automate the tuning and controls of complex accelerators, for example at the Linac Coherent Light Source free-electron laser at SLAC to tune quad\-ru\-pole settings \cite{mcintire2016,duris2020} and at the Central Laser Facility to create the first autonomous laser wakefield accelerator \cite{shalloo2020}. In contrast to these prior studies, which focused on conventional and laser wakefield based acceleration of electron beams, ion beams (in particular, proton-rich isotope beams) need to be controlled in SECAR. Since SECAR is a novel complex device with a lack of previously recorded data (\textit{e.g.}, to train a neural network), online learning, where the model is trained incrementally as it collects individual data instances sequentially from the live separator machine, is required. 

In this article, we demonstrate the first applications of Bayesian optimization with a Gaussian process model to the online beam tuning and ion-optical optimization of the SECAR recoil separator. This method achieved the beam angular deviation and beam width requirements (see Table \ref{table:beam req}) while improving the efficiency of traditional tuning methods and eliminating bias. This work also shows preliminary beam rejection tuning optimization results as part of SECAR commissioning operations. Further tests will be performed in the future.

\subsection{Experimental Setup}\label{setup}

The beamline transport line leading up to the SECAR target includes two pairs of horizontal and vertical electromagnetic steerers upstream of the SECAR target. The two steerer sets are located 7.2 m and 5.4 m upstream of the first SECAR quad\-ru\-pole. These steerers were used in the optimizing of the incoming beam angle and position of the beam at the SECAR target.

The location of the electromagnetic elements from the first quad\-ru\-pole Q1 up to the first mass slits FP2 are shown in Figure \ref{fig:cosy}. Diagnostic elements installed at FP1 and FP2 include Faraday cups, slits, and phosphor coated copper plate viewers. In addition, several Faraday cups are available along the beamline, the first of which, FC1, is installed immediately downstream of Q2. No SECAR target was used during the beam optimizations, however a viewer inside the target chamber was used to check the position and size of the beam spot at the target location. Additionally, a set of current-reading apertures were available in the target chamber, the smallest of which, located directly upstream of the target, has a 6 mm diameter. These apertures limit the incoming beam angle into SECAR to less than 3.4 mrad.

The incoming beam angle and position optimization using the steerers upstream of the SECAR target and the viewer at FP1 is discussed in Section \ref{angle}. Quad\-ru\-poles Q1-Q7 and hex\-a\-poles H1-H2 were included in the optimization of the beam rejection at the mass slits FP2. A detailed discussion is given in Section \ref{optics}. Beam spot location and widths were measured through analysis of FP1 and FP2 viewer images.

\subsection{Traditional Tuning Methods}

Prior to the implementation of Bayesian optimization procedures, current readings in the SECAR beamline were the main diagnostic tools used to optimize the incoming beam angle and position. Once a beam is delivered to FC1, the incoming beam angle and position were optimized by manually minimizing the beam intensity on the target chamber apertures while maximizing the current on the downstream Faraday cup. 
Once a good transmission to the Faraday cup was achieved, the position and size at the viewer in the target chamber were checked. Given the relatively large angular acceptance of the target chamber aperture system (3.4 mrad), adjustments to the required angles below 1 mrad were not possible using this method. 

To optimize the rejection of a separator, methods employed at other facilities such as DRAGON at TRIUMF \cite{Engel2003} and the St. GEORGE recoil separator at Notre Dame University \cite{Couder2008} have typically involved scaling from known tunes that were shown to provide good results in the past, or by manually adjusting quad\-ru\-poles one by one to find the optimal tune about some nominal ion-optical settings obtained from beam physics calculations \cite{Meisel2017}. Since SECAR is a novel device, previous tunes are not readily available. To set the SECAR beamline to a certain tune, the magnetic fields were scaled appropriately according to the magnetic rigidity of the beam from the theoretically optimized COSY INFINITY tune \cite{Berg2018}. These nominal settings require experimental validation and optimization.


\section{Bayesian Optimization}

Bayesian optimization is a gradient-free global maximization or minimization of an unknown black-box function $f$. Each observation of $f$ in the domain is unbiased and possibly noisy, and constitutes a part of prior data collected that informs the decision of where to place the next evaluation (ideally closer to the optimal value) by applying Bayes' theorem. The prediction of where the optimum of $f$ might be in the domain takes into account the model uncertainty when selecting the next point of observation. This method is an iterative search for a better optimum that imposes a probabilistic distribution over $f$. This distribution along with its uncertainty is then used to choose a better optimum, and once the sample of observation is collected at that point, the procedure is repeated. Bayesian optimization depends on two components. The first one is the underlying probabilistic model of the objective function $f$ which describes how it varies with the input parameters (\textit{e.g.}, Gaussian processes), and the second is the choice of acquisition function that places the criterion on how to choose the next point based on the observed data. Refer to \cite{Kushner1964ANM,Mockus1975,brochu2010tutorial} for a thorough introduction to Bayesian optimization. 

The use of Bayesian optimization in the tuning of a complex beamline is in some cases motivated by the time consuming task of operator-dependent system adjustments until a suitable tune is found. In the case of SECAR, manually tuning the four upstream steerers to achieve the beam quality required in Table \ref{table:beam req} at the target location is a time consuming process. Furthermore, the beam rejection from the theoretically optimized COSY tune of SECAR requires validation and optimization. Given the high dimensionality of the system, a robust optimization of the experimental system is needed to validate and improve nominal tune parameters. With a proper model of the beam response in SECAR to changes in beamline parameters (magnet settings in SECAR and upstream), a Bayesian approach presents a good choice for addressing these issues while decreasing the time spent tuning and eliminating the subjective bias introduced by operators.

\subsection{Gaussian Processes}

Gaussian processes (GPs) are popular in Bayesian modeling for regression and classification \cite{Rasmussen2018}. They are easy to implement, flexible, and conveniently provide uncertainty estimates along with their predictions. A GP is an infinite collection of random variables representing the input space (in our case, magnet settings), any finite subset of which is assumed to be jointly Gaussian distributed, and represents the infinite set of possible functions describing the quantity of interest. A GP is a distribution over functions, completely specified by its mean function $\mu$, and covariance function $k$. The covariance function, \textit{i.e.}, the kernel, encodes our assumptions about the objective function and defines the similarities between points in the input space. The choice of covariance function is crucial to the quality of the GP predictions and is directly related to the underlying physics of the beam behavior. As samples are collected, new training data is incorporated into the model, and predictions over the mean $\mu(x)$ and uncertainty $\sigma(x)$ of not yet collected observations are updated. This posterior distribution contains information from both the new data and the prior distribution. For an in-depth review on GPs, refer to \cite{Rasmussen2018}.

\subsection{Covariance Function and Hyperparameters}

A covariance function measures similarities between points in the input space and informs the GP on patterns in the data. In this work, we describe the beam's response to changes in the settings of different magnetic elements by a squared exponential (SE) kernel. With this choice of covariance function, the variance is close to unity for variables with close inputs, and decreases with increasing distance between inputs. The SE function is infinitely differentiable and thus is very smooth, providing a good model for beam response to magnetic field changes. The SE kernel has been successfully used in online optimization at other facilities, \textit{e.g.}, \cite{mcintire2016,duris2020,shalloo2020}. The covariance function is defined as 
\begin{equation}\label{eq:rbf}
    k_{SE}(x_i, x_j) = \sigma^2_{f} \: exp \: (-\frac{|x_i-x_j|^2}{2\ell^{2}}) + \sigma_n^2 \delta_{ij},
\end{equation}
where $\sigma^2_{f}$ is a scaling parameter that represents the observation variance, and $\ell$ is the characteristic lengthscale parameter, a positive constant that roughly describes how an observation informs on neighboring yet unobserved observations. Since experimental data represents a noisy version of the black box function, a Gaussian noise term with variance $\sigma_n$ is added, and $\delta_{ij}$ is the Kronecker delta. Parameters $\sigma_f$, $\ell$ and $\sigma_n$ are known as hyperparameters, and the covariance function learns them empirically from prior observed data by maximizing the log marginal likelihood. A prior of a Gaussian form was initialized for the lengthscale $\ell$ and noise $\sigma_n$ hyperparameters with a mean and variance based on empirical data collected during commissioning, typically $\ell \sim \mathcal{N}(2, 1)$ (A) and $\sigma_n^2 \sim \mathcal{N}(1, 0.5)$ (pixels$^2$). As the sampled data was iteratively collected, the hyperparameters were optimized and the priors were updated by maximizing the probability of the model given the data. 

\subsection{Acquisition Function}

Bayesian optimization uses the probability distribution constructed from observed data to decide where to evaluate the objective function at the next step guided by an acquisition function. In this work we selected a lower confidence bound (LCB) acquisition function \cite{Cox97} constructed from the GP posterior mean function $\mu$(x) and its uncertainty $\sigma$(x)
\begin{equation}\label{eq:lcb}
LCB(x) = \mu(x) - \xi \, \sigma(x),
\end{equation}
where $\xi$ is the user defined exploration weight that directly balances the trade-off between exploiting regions of low mean (since we are searching for the minimum) and exploring regions of large uncertainty. The LCB was minimized and the next measurement was chosen to be sampled at that minimum and added to the GP.

\section{SECAR Beam Commissioning}

SECAR beam commissioning described in this work has been performed using \textsuperscript{2}H\textsuperscript{1+}, \textsuperscript{133}Cs\textsuperscript{41+} and \textsuperscript{20}Ne\textsuperscript{8+} beams spanning a magnetic rigidity range of 0.1444 to 0.4667 Tm.

The online Bayesian optimization developed during commissioning used the Python GPy library \cite{GPy}, along with the associated GPyOpt tool \cite{Authors2016} for the GP framework. The program was integrated into the SECAR magnet and diagnostics control system using PyEpics, a Python interface to the EPICS Channel Access (CA) library for the EPICS control system \cite{Newville_2017}.


\section{Incoming Beam Angle and Position Optimization} \label{angle}

In the absence of a direct way to measure and optimize the incoming beam angle at the SECAR target, an indirect method was adopted using existing beamline elements to ensure the incoming beam is optimized with respect to the SECAR ion-optical axis. Here, we describe the method that was used with the first two quad\-ru\-poles, Q1 and Q2, and the viewer at FP1. The same method applies for any suitable combination of quad\-ru\-poles and downstream viewers along the SECAR beamline. This was especially helpful to adjust small angular deviations whose effects were negligible at upstream viewer locations but significant towards the last sections of SECAR.

\subsection{Method}

When the incoming beam is aligned with the central axis of a quad\-ru\-pole magnet, changes in the quad\-ru\-pole strength focus and de-focus the beam. If the path of the beam deviates from the ion-optical axis, the beam experiences imbalanced forces from the quad\-ru\-pole fields and steers in the unbalanced direction proportionally to the quad\-ru\-pole excitation. Optimization of the incoming beam angle and position at the SECAR target was accomplished by minimizing the steering produced by SECAR quad\-ru\-poles downstream of the target. The steering due to the misalignment of the beam with respect to the optical axis of the SECAR quad\-ru\-poles was observed and measured using digitized images from diagnostic viewers along the SECAR beamline. The incoming beam angle and position were controlled using the two sets of steerers upstream of the target described in Section \ref{setup}. 

The amount of steering produced by the SECAR quad\-ru\-poles was defined as the average beam movement observed between four different quad\-ru\-poles settings. The four quad\-ru\-pole settings were arbitrarily chosen to provide beam spot effects that were visibly different enough for analysis. The center location of each beam spot was measured (X and Y) from the digitized viewer image, and the mean distance between the four center locations was calculated. 

The algorithm to minimize the beam steering is described in Algorithm \ref{bayes algo}. The criterion for stopping the optimization was the convergence of the algorithm to a minimum in the beam movement function. Once the minimum in the objective function was found, the conditions of a centered beam at the target and full transmission through SECAR where verified manually using the viewer at the target and Faraday cups along the beamline. 
\begin{algorithm}[H]
\caption{Bayesian optimization for SECAR} \label{bayes algo}
\begin{algorithmic}[1]
\State Randomly select initial steerer settings and evaluate average steering function $f$ at viewer 
\While {criterion not met} 
    \State Compute $x^*$ =  argmin (LCB($x$)) 
    \State Set steerer settings to new $x^*$ 
    \State Evaluate $f$ to get average beam movement at viewer 
    \State Add new observation to set of samples 
\EndWhile 
\State Set $x^*$ to the final tune 
\end{algorithmic}
\end{algorithm}

Following this method, all four upstream steerer settings were optimized such that the incoming beam angle was minimized entering SECAR, the beam spot on target was minimized, and the beam aligned with the SECAR magnetic beam axis. In cases where fine tuning was needed in one direction (only horizontal or only vertical steering), the two corresponding (either horizontal or vertical) steerers were optimized while the other two were kept fixed. This method was applied at every viewer up to the final detector plane along the SECAR beamline utilizing the quad\-ru\-poles upstream of each viewer to probe and fine tune the angular deviation at each location. 

\subsection{Results}

\begin{figure}
    \centering
    \includegraphics[width=6.5cm]{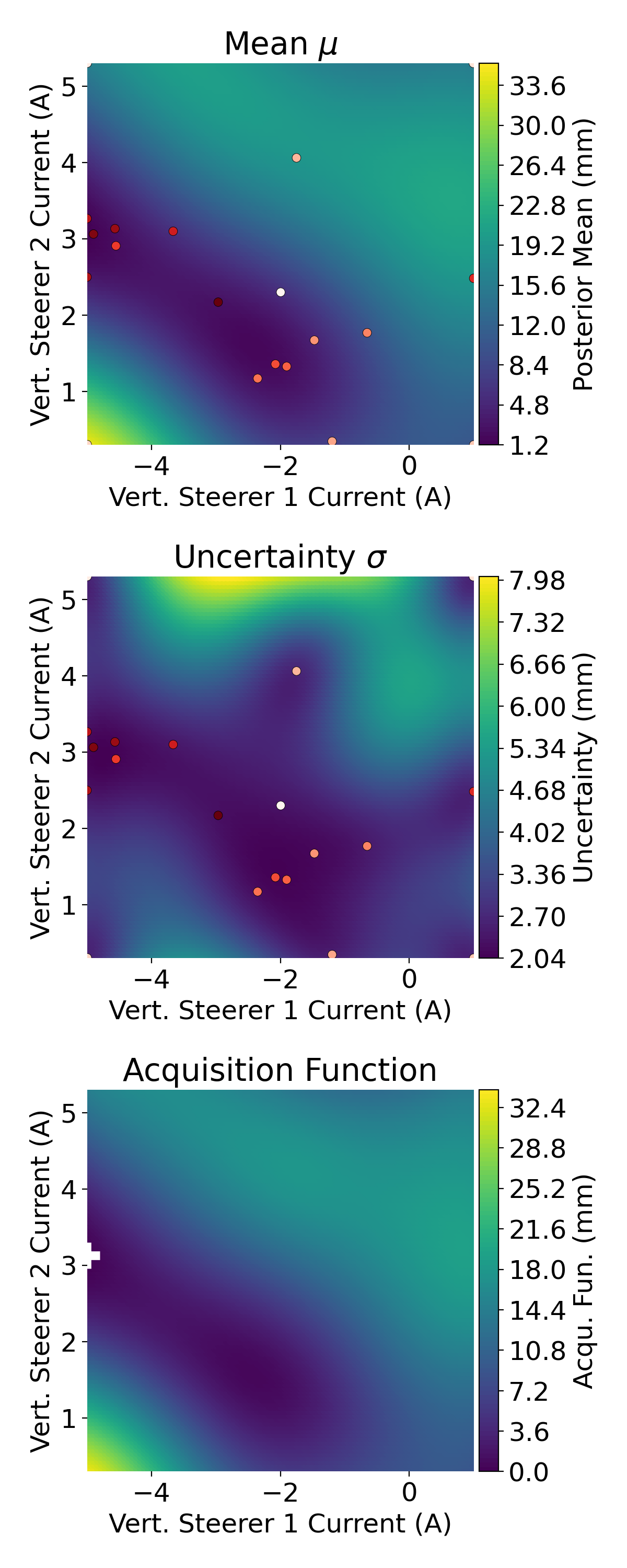}
    \caption{Mean $\mu$ (a), standard deviation $\sigma$ (b) and acquisition function (c) for a 2D steerer optimization run. The observations are shown in (a) and (b) by dots shaded by time of observation, with the darkest shade being the most recent observation. The next sample point is shown in (c) by the white cross indicating the minimum of the LCB acquisition function.}
    \label{fig:2d results}
\end{figure}

Figure \ref{fig:2d results} presents the mean, standard deviation, and LCB acquisition function evolution as observations were sampled for a typical 2D vertical optimization. 
Observations are shown as circular data points (starting with white, they get darker up to the most recent observation in dark red) overlayed on the mean (a) and uncertainty standard deviation (b) plots.  On the acquisition function plot, the white cross indicates the next point to be sampled, \textit{i.e.}, the minimum of the acquisition function. The LCB exploration weight $\xi$ was constant at 0.5. Hyperparameter priors were selected as $\ell_{prior} = \mathcal{N}(2, 1)$ (A) and $\sigma_{n, prior}^2 = \mathcal{N}(3.5, 0.5)$ (pixels$^2$), and were updated at each iteration. Their mean values at the last observation were $\ell$ = 1.2 A for the lengthscale and $\sigma_{n}$ = 1.9 pixels for the noise. Both hyperparameters varied between tunes and at different locations in the beamline, making an empirical optimization using prior data collection the best option currently available to us. 

\begin{figure}
    \centering
    \includegraphics[width=8.6cm]{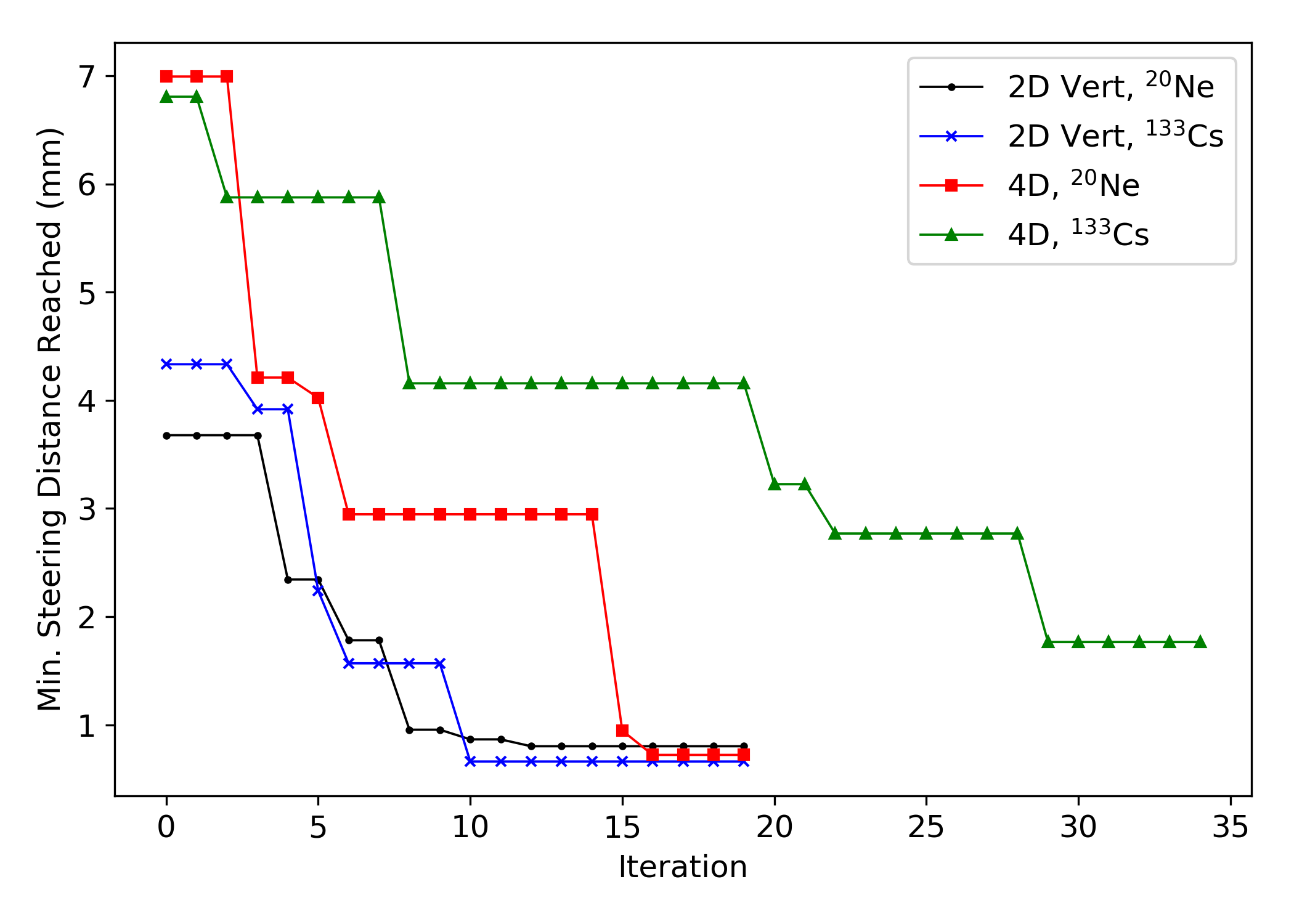}
    \caption{Best steering distance reached as a function of number of GP iterations, for 2D and 4D optimizations.}
    \label{fig:steererconv}
\end{figure}

Figure \ref{fig:steererconv} shows the typical number of iterations to reach convergence for 2D and 4D steerer setting optimizations with different beams.
In general, all 2D optimizations converged within 15 to 20 iterations, while 4D optimizations generally took about double the number of iterations, occasionally needing around 60 when the tune was particularly difficult. At each iteration, adjusting the quad\-ru\-poles takes up to 10 s until they settle. Thus, typical optimization times range between 20 to 40 minutes, which is a significant improvement over the total time spent manually tuning (at least 1-2 hours). 

\begin{figure}
    \centering
    \includegraphics[width=8.6cm]{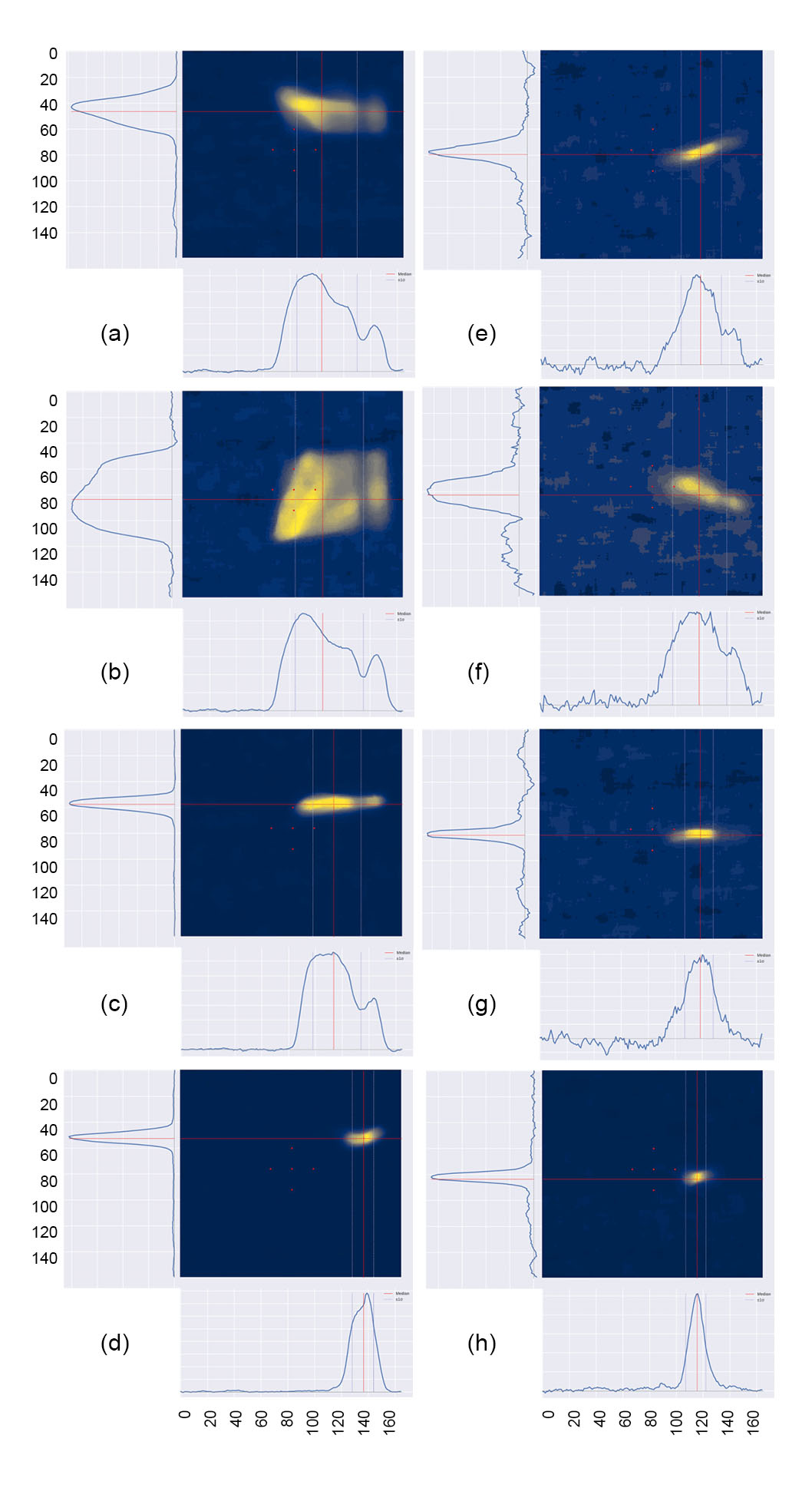}
    \caption{Beam spot seen at FP1 before (a-d) and after (e-h) steerers optimization with a Ne beam. Beam movement due to unbalanced quadrupole forces decreased from an average of 7 mm to an average of 0.2 mm. All axes are in pixels, and 16 pixels correspond to 5 mm.}
    \label{fig:steering results}
\end{figure}

Figure \ref{fig:steering results} a-d (e-f) show the beam at the FP1 viewer for each quad\-ru\-pole setting before (after) a 4D optimization.
All images were taken at the charge selection FP1 before the bending B1-B2 dipole magnets were fine-tuned, so the x-position is not the final beam position. Only the first two quads upstream of the first two bending magnets, Q1 and Q2, were varied in strength in each image to create a deflection due to the incoming beam angle while Q3-Q5 were set to zero. These results were obtained after 16 iterations. The average beam movement between the four quad\-ru\-pole settings was reduced from 7 mm to 0.2 mm after steering settings optimization.  

To quantify the incoming beam angle, the COSY INFINITY model was used to approximate the incoming angle that could create the observed level of steering movement. This was achieved by comparing beam spot locations in COSY to those obtained experimentally with each quad\-ru\-pole tune used to calculate the steering. A $\chi^2$-minimization was performed to find the incoming beam angle that would best match the beam movement due to different quad settings. For the example shown in Figure \ref{fig:steering results}, the $\chi^2$-minimization found that the initial angular deviation that could lead to a 7 mm steering is 1.6 mrad, and after optimization, the angle is reduced to 0.8 mrad. 

\section{Beam Rejection Optimization} \label{optics}

As discussed in Section \ref{intro}, the beam spot width needs to be minimized at the mass selection focal planes to obtain maximal rejection of unreacted beam particles. Preliminary Bayesian optimization to minimize the b\-eam spot width at FP2 while maintaining the energy acceptance of the system is discussed in the next sections. 

\subsection{Method}

The optimization was initialized to the nominal COSY INFINITY magnet settings, with all seven quad\-ru\-poles and up to two hex\-a\-poles upstream of the mass selection FP2
considered as free parameters. The unknown function $f$ to be minimized as a function of the magnet strength parameters corresponds to the beam spot width, defined as the median $\pm1\sigma$. In this work, the optimization focused on the edges of the energy acceptance at $\pm$4\% $dE/E$ and checked that the focused beam spot position did not move significantly within that energy range. In the future, similar measurements need to be performed for the angular acceptance (see \cite{Berg2018} for more details on the SECAR design parameters). To ensure full transmission along the beamline, an additional constraint on the width of the beam spot in the y-direction was added to the objective function. The function $f$ at iteration $i$ took the form
\begin{equation}\label{eq: f optics}
    f = (\frac{x_i+ w^x_i - x_{init}}{20})^4 + (\frac{w^y_i}{70})^4,
\end{equation}
where $x_i$ is the position of the beam at the mass selection FP2 viewer at iteration $i$, $w^{x}_i$ and $w^{y}_i$ are the beam spot widths in the x- and y-directions respectively at iteration $i$, and $x_{init}$ is the position of the SECAR beam axis. The function form and weights were derived empirically after several tests runs to maximize transmission and to minimize beam movement away from the beam axis. With the updated objective function, Algorithm \ref{bayes algo} was applied with the following changes: $f$ was sampled following Equation \ref{eq: f optics}, and the inputs $x$ were in this case the SECAR beamline multipole settings being optimized. Bayesian optimization was done at 0\%, +4\% and -4\% $dE/E$ for several beams and energies. The magnet settings providing the smallest beam widths were selected as the final optimal tune for each magnetic rigidity. 

\subsection{Results}

This method was tested during SECAR commissioning runs with a \textsuperscript{133}Cs\textsuperscript{41+} beam at 1 MeV/nucleon. In this run, the quad\-ru\-poles Q1 to Q7 and the hex\-a\-poles Hex1 and Hex2 were optimized. A second optimization was run with a \textsuperscript{20}Ne\textsuperscript{8+} beam at 1.8 MeV/nucleon where Q1 to Q7 were optimized, while Hex1 and Hex2 were kept a zero excitation. It should be noted that the Q1 magnet has a
combined design where an adjustable hex\-a\-pole magnet (Q1Hex) along with a correcting dipole
(Q1Dipole) component can be set separately. The Q1Hex component (but not the Q1Dipole component) was included in the second optimization as a parameter, making it an 8-dimensional optimization.

\begin{figure}
    \centering
    \includegraphics[width=8.6cm]{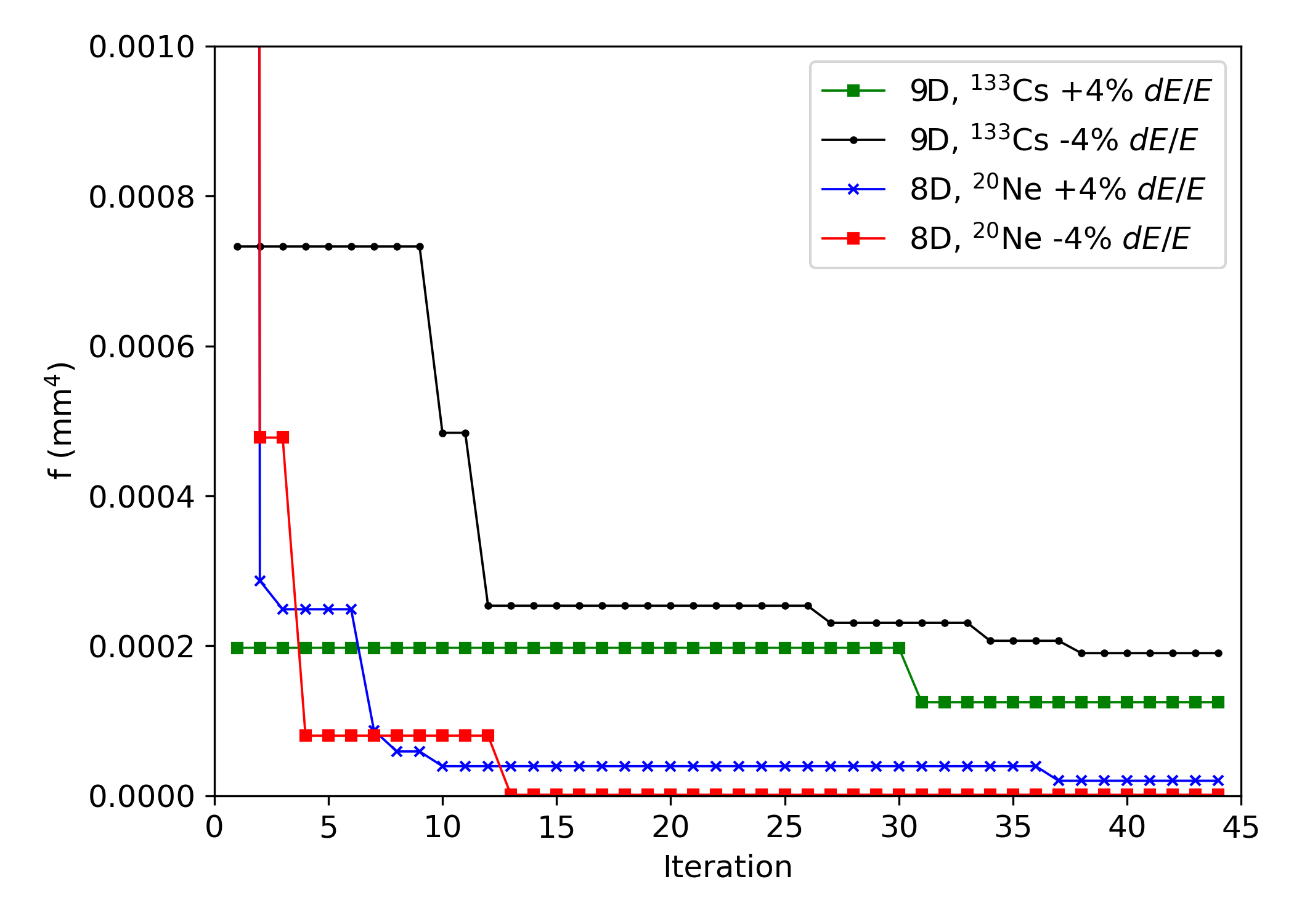}
    \caption{Objective function defined in Equation \ref{eq: f optics} reached as a function of number of GP iterations, for 8D (seven quad\-ru\-pole and one hex\-a\-pole magnets) and 9D (seven quad\-ru\-pole and two hex\-a\-pole magnets) optimizations within $\pm$4\% $dE/E$.}
    \label{fig:opticsconv}
\end{figure}
Figure \ref{fig:opticsconv} shows the speed of convergence for the optimization runs within a $\pm$4\% $dE/E$ acceptance. The GP optimization ran for approximately 1.5 hours at each energy. Scaling the beamline (including the velocity filter) $\pm$4\% in energy was done manually, making the total beam rejection optimization time approximately 5 hours. While we have not attempted manual tuning of the quadrupoles to optimize the rejection in SECAR, we estimate it would take between 8 and 16 hours based on magnet optimization in other recoil separators with a similar number of magnets, such as St. GEORGE \cite{Meisel2017}.

Once the minimum of $f$ had been found, the beam size width defined as $\pm2\sigma$ of the beam intensity for a beam with a $\pm4$\% energy spread was measured for the COSY INFINITY nominal tune and for the GP best tune. The final results are summarized in Table \ref{tab:slit positions} for each beam. 

\begin{figure}
    \centering
    \includegraphics[width=8.6cm]{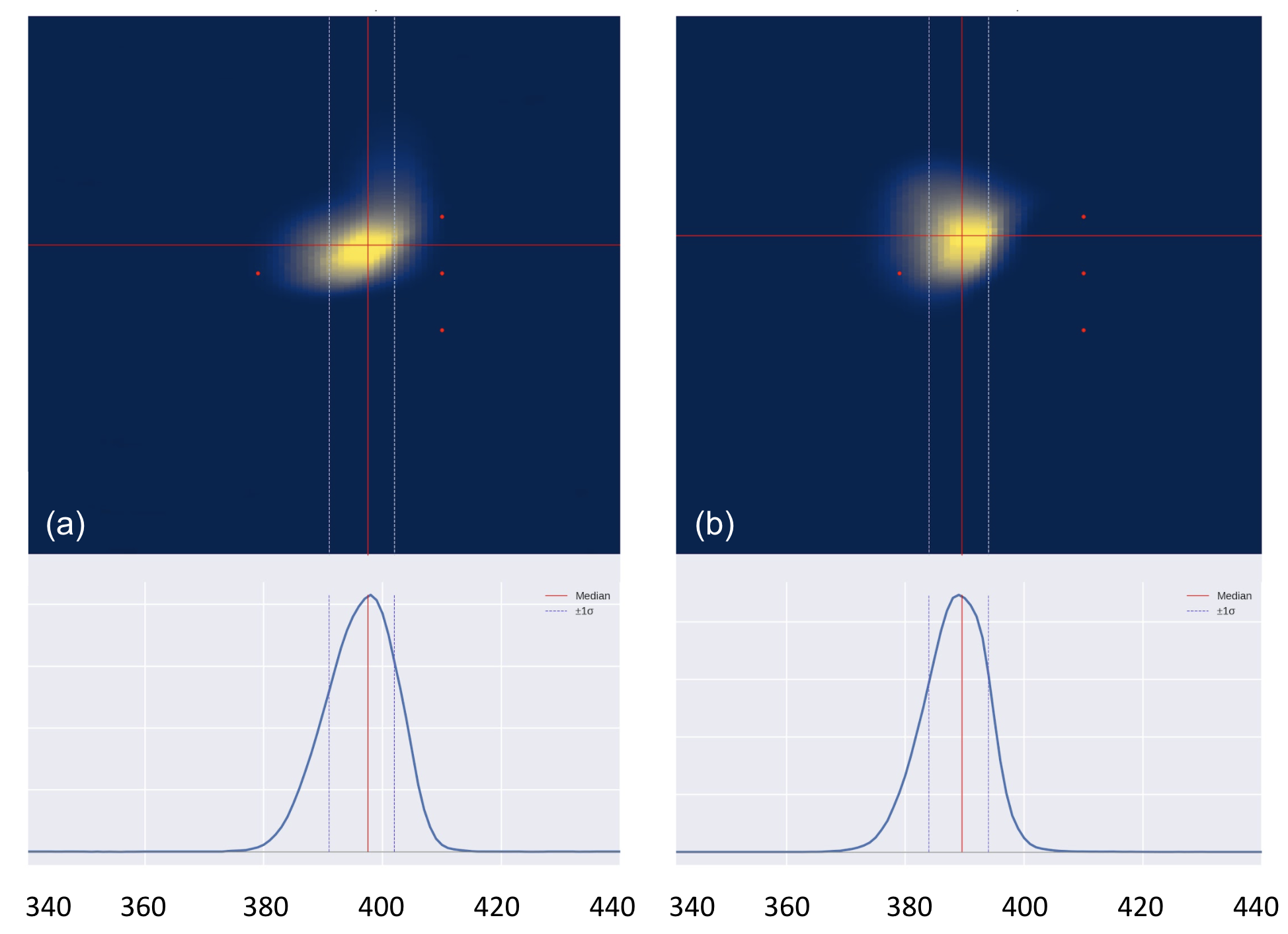}
    \caption{(a) Nominal and (b) after the GP optimization of the width at FP2 for the \textsuperscript{133}Cs commissioning run (see Table \ref{tab:slit positions}). Red dots indicate 5 mm. X-axis scale is in pixels, where 31 pixels is 5 mm.}
    \label{fig:width cs}
\end{figure}

\begin{table}
    \centering
    \begin{tabular}{c c c c}
    \hline
    \textbf{Beam} & \textbf{B$\rho$ (Tm)} & \textbf{Nom. Gap (mm)}	& \textbf{GP Gap (mm)} \\
    \hline
    \textsuperscript{133}Cs  &	  0.4667   &  6.13 &  4.19 \\ [0.5ex]
    \textsuperscript{20}Ne   &   0.3887   &  8.23 &  12.26 \\ [0.5ex]
    \end{tabular}
    \caption{Slit gaps needed to admit $\pm2\sigma$ of the recoils within a $\pm$4\% $dE/E$ acceptance for the nominal tunes and the final GP tunes.}
    \label{tab:slit positions}
\end{table}
A 32\% decrease in the beam spot width was achieved with the Cs beam following this procedure. A comparison of the beam spot before and after optimization is given in Figure \ref{fig:width cs}. The total shift in x-position between the two images is about 1.3 mm and no significant change in the length in the y-direction is seen.

No improvement was achieved for the Ne beam compared to the COSY nominal settings. This may be attributed to the inclusion of Q1Hex in the optimization, which may have steering effects
on the beam that were not compensated by the Q1Dipole, possibly driving the optimization
away from the best optics. These results also indicate that it is crucial to optimize Hex1 and Hex2 and correcting higher order contributions to achieve the desired rejection.

\section{Summary and Outlook}

In this article, we presented the first development of online Bayesian optimization with Gaussian processes for tuning an ion beam in a nuclear astrophysics recoil separator. We showed the method increases the efficiency and precision in achieving the stringent requirements needed for optimal separator performance (summarized in Table \ref{table:beam req}) as compared to traditional manual tuning methods.

Our Bayesian optimization algorithm was shown to minimize the incoming beam angular deviation within the specified requirements down to angles below 1 mrad. This method can be used for all separator tuning, and can generally be applied to other similar beamlines. Bayesian optimization of the quad\-ru\-pole and hex\-a\-pole magnets was performed within an upper limit of the energy acceptance of $\pm$4\%. We showed that our Bayesian method validated the nominal COSY INFINITY model settings, and found improved settings that decreased the beam size at the mass slits by 32\%  as seen in the presented case of the \textsuperscript{133}Cs beam. This optimization can be run ahead of each scientific experiment in order to provide the best possible beam suppression.

As this is an initial implementation of this algorithm tested during SECAR commissioning operations, several improvements are being explored. For instance, the extensive database of accelerator historical data as well as physics simulations of SECAR and the accelerator may be used in implementing physics-informed optimizations of incoming beam parameters \cite{hanuka2021}. Additionally, beam specific priors can be developed by establishing a relationship between beam species, beam rigidity, and the GP kernel hyperparameters. Future work can also include thorough ion-optical comparisons of the beam angle at different foci in SECAR after the beamline is fully tuned and at different stages of the optimization to gain a better understanding of the system. 

As quad\-ru\-poles are designed to work in pairs or triplets, a physics-informed correlation matrix can be added to the covariance function when optimizing the beam spot width to arrive at better optics more efficiently. This can be obtained using the currently available COSY INFINITY model of SECAR. Moreover, the algorithm can be expanded into a multi-objective Bayesian optimization \cite{roussel2020multiobjective} to improve the COSY model in achieving a high mass resolution for different scientific experiments. By simultaneously optimizing several transfer matrix elements including the distance from the beam center in the x-direction given the initial fractional angle, energy and mass difference of the beam particles compared to mean value at the mass slits, the full Pareto front of this optimization can be reached with fewer observations. The Pareto front is defined as a set of points that optimally balances the trade-offs between multiple competing objectives simultaneously. This multi-objective approach can also be useful in online implementations such as the ones presented in this work where multiple optimization goals are often being sought out. 

\begin{acknowledgments}
This material is based upon work supported by the U.S. Department of Energy, Office of Science, Office of Nuclear Physics under award number DE-SC0014384, and the U.S. National Science Foundation under award numbers PHY-1624942, PHY-1913554, PHY-1102511, and PHY 14-30152 (JINA-CEE).
\end{acknowledgments}


\begin{thebibliography}{21}%
\makeatletter
\providecommand \@ifxundefined [1]{%
 \@ifx{#1\undefined}
}%
\providecommand \@ifnum [1]{%
 \ifnum #1\expandafter \@firstoftwo
 \else \expandafter \@secondoftwo
 \fi
}%
\providecommand \@ifx [1]{%
 \ifx #1\expandafter \@firstoftwo
 \else \expandafter \@secondoftwo
 \fi
}%
\providecommand \natexlab [1]{#1}%
\providecommand \enquote  [1]{``#1''}%
\providecommand \bibnamefont  [1]{#1}%
\providecommand \bibfnamefont [1]{#1}%
\providecommand \citenamefont [1]{#1}%
\providecommand \href@noop [0]{\@secondoftwo}%
\providecommand \href [0]{\begingroup \@sanitize@url \@href}%
\providecommand \@href[1]{\@@startlink{#1}\@@href}%
\providecommand \@@href[1]{\endgroup#1\@@endlink}%
\providecommand \@sanitize@url [0]{\catcode `\\12\catcode `\$12\catcode
  `\&12\catcode `\#12\catcode `\^12\catcode `\_12\catcode `\%12\relax}%
\providecommand \@@startlink[1]{}%
\providecommand \@@endlink[0]{}%
\providecommand \url  [0]{\begingroup\@sanitize@url \@url }%
\providecommand \@url [1]{\endgroup\@href {#1}{\urlprefix }}%
\providecommand \urlprefix  [0]{URL }%
\providecommand \Eprint [0]{\href }%
\providecommand \doibase [0]{https://doi.org/}%
\providecommand \selectlanguage [0]{\@gobble}%
\providecommand \bibinfo  [0]{\@secondoftwo}%
\providecommand \bibfield  [0]{\@secondoftwo}%
\providecommand \translation [1]{[#1]}%
\providecommand \BibitemOpen [0]{}%
\providecommand \bibitemStop [0]{}%
\providecommand \bibitemNoStop [0]{.\EOS\space}%
\providecommand \EOS [0]{\spacefactor3000\relax}%
\providecommand \BibitemShut  [1]{\csname bibitem#1\endcsname}%
\let\auto@bib@innerbib\@empty
\bibitem [{\citenamefont {Kester}\ \emph {et~al.}(2011)\citenamefont {Kester}
  \emph {et~al.}}]{Kester2011}%
  \BibitemOpen
  \bibfield  {author} {\bibinfo {author} {\bibfnamefont {O.}~\bibnamefont
  {Kester}} \emph {et~al.},\ }\bibfield  {title} {\bibinfo {title} {{ReA3 - the
  Rare Isotope Re-accelerator at MSU}},\ }in\ \href@noop {} {\emph {\bibinfo
  {booktitle} {{25th International Linear Accelerator Conference}}}}\ (\bibinfo
  {year} {2011})\ p.\ \bibinfo {pages} {MO203}\BibitemShut {NoStop}%
\bibitem [{\citenamefont {{Berg}}\ \emph {et~al.}(2018)\citenamefont {{Berg}},
  \citenamefont {{Couder}}, \citenamefont {{Moran}}, \citenamefont {{Smith}},
  \citenamefont {{Wiescher}}, \citenamefont {{Schatz}}, \citenamefont
  {{Hager}}, \citenamefont {{Wrede}}, \citenamefont {{Montes}}, \citenamefont
  {{Perdikakis}}, \citenamefont {{Wu}}, \citenamefont {{Zeller}}, \citenamefont
  {{Smith}}, \citenamefont {{Bardayan}}, \citenamefont {{Chipps}},
  \citenamefont {{Pain}}, \citenamefont {{Blackmon}}, \citenamefont {{Greife}},
  \citenamefont {{Rehm}},\ and\ \citenamefont {{Janssens}}}]{Berg2018}%
  \BibitemOpen
  \bibfield  {author} {\bibinfo {author} {\bibfnamefont {G.~P.~A.}\
  \bibnamefont {{Berg}}}, \bibinfo {author} {\bibfnamefont {M.}~\bibnamefont
  {{Couder}}}, \bibinfo {author} {\bibfnamefont {M.~T.}\ \bibnamefont
  {{Moran}}}, \bibinfo {author} {\bibfnamefont {K.}~\bibnamefont {{Smith}}},
  \bibinfo {author} {\bibfnamefont {M.}~\bibnamefont {{Wiescher}}}, \bibinfo
  {author} {\bibfnamefont {H.}~\bibnamefont {{Schatz}}}, \bibinfo {author}
  {\bibfnamefont {U.}~\bibnamefont {{Hager}}}, \bibinfo {author} {\bibfnamefont
  {C.}~\bibnamefont {{Wrede}}}, \bibinfo {author} {\bibfnamefont
  {F.}~\bibnamefont {{Montes}}}, \bibinfo {author} {\bibfnamefont
  {G.}~\bibnamefont {{Perdikakis}}}, \bibinfo {author} {\bibfnamefont
  {X.}~\bibnamefont {{Wu}}}, \bibinfo {author} {\bibfnamefont {A.}~\bibnamefont
  {{Zeller}}}, \bibinfo {author} {\bibfnamefont {M.~S.}\ \bibnamefont
  {{Smith}}}, \bibinfo {author} {\bibfnamefont {D.~W.}\ \bibnamefont
  {{Bardayan}}}, \bibinfo {author} {\bibfnamefont {K.~A.}\ \bibnamefont
  {{Chipps}}}, \bibinfo {author} {\bibfnamefont {S.~D.}\ \bibnamefont
  {{Pain}}}, \bibinfo {author} {\bibfnamefont {J.}~\bibnamefont {{Blackmon}}},
  \bibinfo {author} {\bibfnamefont {U.}~\bibnamefont {{Greife}}}, \bibinfo
  {author} {\bibfnamefont {K.~E.}\ \bibnamefont {{Rehm}}},\ and\ \bibinfo
  {author} {\bibfnamefont {R.~V.~F.}\ \bibnamefont {{Janssens}}},\ }\bibfield
  {title} {\bibinfo {title} {{Design of SECAR a recoil mass separator for
  astrophysical capture reactions with radioactive beams}},\ }\href
  {https://doi.org/10.1016/j.nima.2017.08.048} {\bibfield  {journal} {\bibinfo
  {journal} {Nucl. Instrum. Methods Phys. Res. A}\ }\textbf {\bibinfo {volume}
  {877}},\ \bibinfo {pages} {87} (\bibinfo {year} {2018})}\BibitemShut
  {NoStop}%
\bibitem [{\citenamefont {Berz}()}]{MAKINO2006346}%
  \BibitemOpen
  \bibfield  {author} {\bibinfo {author} {\bibfnamefont {M.}~\bibnamefont
  {Berz}},\ }\href@noop {} {\bibinfo {title} {Cosy infinity}},\ \bibinfo
  {howpublished} {\url{http://www.cosyinfinity.org/}}\BibitemShut {NoStop}%
\bibitem [{\citenamefont {Chipps}\ \emph {et~al.}(2014)\citenamefont {Chipps},
  \citenamefont {Greife}, \citenamefont {Bardayan}, \citenamefont {Blackmon},
  \citenamefont {Kontos}, \citenamefont {Linhardt}, \citenamefont {Matos},
  \citenamefont {Pain}, \citenamefont {Pittman}, \citenamefont {Sachs},
  \citenamefont {Schatz}, \citenamefont {Schmitt}, \citenamefont {Smith},\ and\
  \citenamefont {Thompson}}]{CHIPPS2014}%
  \BibitemOpen
  \bibfield  {author} {\bibinfo {author} {\bibfnamefont {K.}~\bibnamefont
  {Chipps}}, \bibinfo {author} {\bibfnamefont {U.}~\bibnamefont {Greife}},
  \bibinfo {author} {\bibfnamefont {D.}~\bibnamefont {Bardayan}}, \bibinfo
  {author} {\bibfnamefont {J.}~\bibnamefont {Blackmon}}, \bibinfo {author}
  {\bibfnamefont {A.}~\bibnamefont {Kontos}}, \bibinfo {author} {\bibfnamefont
  {L.}~\bibnamefont {Linhardt}}, \bibinfo {author} {\bibfnamefont
  {M.}~\bibnamefont {Matos}}, \bibinfo {author} {\bibfnamefont
  {S.}~\bibnamefont {Pain}}, \bibinfo {author} {\bibfnamefont {S.}~\bibnamefont
  {Pittman}}, \bibinfo {author} {\bibfnamefont {A.}~\bibnamefont {Sachs}},
  \bibinfo {author} {\bibfnamefont {H.}~\bibnamefont {Schatz}}, \bibinfo
  {author} {\bibfnamefont {K.}~\bibnamefont {Schmitt}}, \bibinfo {author}
  {\bibfnamefont {M.}~\bibnamefont {Smith}},\ and\ \bibinfo {author}
  {\bibfnamefont {P.}~\bibnamefont {Thompson}},\ }\bibfield  {title} {\bibinfo
  {title} {The jet experiments in nuclear structure and astrophysics (jensa)
  gas jet target},\ }\href
  {https://doi.org/https://doi.org/10.1016/j.nima.2014.06.042} {\bibfield
  {journal} {\bibinfo  {journal} {Nucl. Instrum. Methods Phys. Res. A}\
  }\textbf {\bibinfo {volume} {763}},\ \bibinfo {pages} {553} (\bibinfo {year}
  {2014})}\BibitemShut {NoStop}%
\bibitem [{\citenamefont {Schmidt}\ \emph {et~al.}(2018)\citenamefont
  {Schmidt}, \citenamefont {Chipps}, \citenamefont {Ahn}, \citenamefont
  {Bardayan}, \citenamefont {Browne}, \citenamefont {Greife}, \citenamefont
  {Meisel}, \citenamefont {Montes}, \citenamefont {O’Malley}, \citenamefont
  {Ong}, \citenamefont {Pain}, \citenamefont {Schatz}, \citenamefont {Smith},
  \citenamefont {Smith},\ and\ \citenamefont {Thompson}}]{SCHMIDT2018}%
  \BibitemOpen
  \bibfield  {author} {\bibinfo {author} {\bibfnamefont {K.}~\bibnamefont
  {Schmidt}}, \bibinfo {author} {\bibfnamefont {K.}~\bibnamefont {Chipps}},
  \bibinfo {author} {\bibfnamefont {S.}~\bibnamefont {Ahn}}, \bibinfo {author}
  {\bibfnamefont {D.}~\bibnamefont {Bardayan}}, \bibinfo {author}
  {\bibfnamefont {J.}~\bibnamefont {Browne}}, \bibinfo {author} {\bibfnamefont
  {U.}~\bibnamefont {Greife}}, \bibinfo {author} {\bibfnamefont
  {Z.}~\bibnamefont {Meisel}}, \bibinfo {author} {\bibfnamefont
  {F.}~\bibnamefont {Montes}}, \bibinfo {author} {\bibfnamefont
  {P.}~\bibnamefont {O’Malley}}, \bibinfo {author} {\bibfnamefont {W.-J.}\
  \bibnamefont {Ong}}, \bibinfo {author} {\bibfnamefont {S.}~\bibnamefont
  {Pain}}, \bibinfo {author} {\bibfnamefont {H.}~\bibnamefont {Schatz}},
  \bibinfo {author} {\bibfnamefont {K.}~\bibnamefont {Smith}}, \bibinfo
  {author} {\bibfnamefont {M.}~\bibnamefont {Smith}},\ and\ \bibinfo {author}
  {\bibfnamefont {P.}~\bibnamefont {Thompson}},\ }\bibfield  {title} {\bibinfo
  {title} {Status of the jensa gas-jet target for experiments with rare isotope
  beams},\ }\href {https://doi.org/https://doi.org/10.1016/j.nima.2018.09.052}
  {\bibfield  {journal} {\bibinfo  {journal} {Nucl. Instrum. Methods Phys. Res.
  A}\ }\textbf {\bibinfo {volume} {911}},\ \bibinfo {pages} {1} (\bibinfo
  {year} {2018})}\BibitemShut {NoStop}%
\bibitem [{\citenamefont {{M. McIntire, T. Cope, S. Ermon, and
  others}}(2016)}]{mcintire2016}%
  \BibitemOpen
  \bibfield  {author} {\bibinfo {author} {\bibnamefont {{M. McIntire, T. Cope,
  S. Ermon, and others}}},\ }\bibfield  {title} {\bibinfo {title} {{Bayesian
  Optimization of FEL Performance at LCLS}},\ }in\ \href@noop {} {\emph
  {\bibinfo {booktitle} {{Proceedings of the 7th International Particle
  Accelerator Conference (IPAC)}}}}\ (\bibinfo {year} {2016})\ p.\ \bibinfo
  {pages} {WEPOW055}\BibitemShut {NoStop}%
\bibitem [{\citenamefont {Duris}\ \emph {et~al.}(2020)\citenamefont {Duris},
  \citenamefont {Kennedy}, \citenamefont {Hanuka}, \citenamefont {Shtalenkova},
  \citenamefont {Edelen}, \citenamefont {Baxevanis}, \citenamefont {Egger},
  \citenamefont {Cope}, \citenamefont {McIntire}, \citenamefont {Ermon},\ and\
  \citenamefont {Ratner}}]{duris2020}%
  \BibitemOpen
  \bibfield  {author} {\bibinfo {author} {\bibfnamefont {J.}~\bibnamefont
  {Duris}}, \bibinfo {author} {\bibfnamefont {D.}~\bibnamefont {Kennedy}},
  \bibinfo {author} {\bibfnamefont {A.}~\bibnamefont {Hanuka}}, \bibinfo
  {author} {\bibfnamefont {J.}~\bibnamefont {Shtalenkova}}, \bibinfo {author}
  {\bibfnamefont {A.}~\bibnamefont {Edelen}}, \bibinfo {author} {\bibfnamefont
  {P.}~\bibnamefont {Baxevanis}}, \bibinfo {author} {\bibfnamefont
  {A.}~\bibnamefont {Egger}}, \bibinfo {author} {\bibfnamefont
  {T.}~\bibnamefont {Cope}}, \bibinfo {author} {\bibfnamefont {M.}~\bibnamefont
  {McIntire}}, \bibinfo {author} {\bibfnamefont {S.}~\bibnamefont {Ermon}},\
  and\ \bibinfo {author} {\bibfnamefont {D.}~\bibnamefont {Ratner}},\
  }\bibfield  {title} {\bibinfo {title} {Bayesian optimization of a
  free-electron laser},\ }\href
  {https://doi.org/10.1103/PhysRevLett.124.124801} {\bibfield  {journal}
  {\bibinfo  {journal} {Phys. Rev. Lett.}\ }\textbf {\bibinfo {volume} {124}},\
  \bibinfo {pages} {124801} (\bibinfo {year} {2020})}\BibitemShut {NoStop}%
\bibitem [{\citenamefont {Shalloo}\ \emph {et~al.}(2020)\citenamefont
  {Shalloo}, \citenamefont {Dann}, \citenamefont {Gruse}, \citenamefont
  {Underwood}, \citenamefont {Antoine}, \citenamefont {Arran}, \citenamefont
  {Backhouse}, \citenamefont {Baird}, \citenamefont {Balcazar}, \citenamefont
  {Bourgeois},\ and\ \citenamefont {et~al.}}]{shalloo2020}%
  \BibitemOpen
  \bibfield  {author} {\bibinfo {author} {\bibfnamefont {R.~J.}\ \bibnamefont
  {Shalloo}}, \bibinfo {author} {\bibfnamefont {S.~J.~D.}\ \bibnamefont
  {Dann}}, \bibinfo {author} {\bibfnamefont {J.-N.}\ \bibnamefont {Gruse}},
  \bibinfo {author} {\bibfnamefont {C.~I.~D.}\ \bibnamefont {Underwood}},
  \bibinfo {author} {\bibfnamefont {A.~F.}\ \bibnamefont {Antoine}}, \bibinfo
  {author} {\bibfnamefont {C.}~\bibnamefont {Arran}}, \bibinfo {author}
  {\bibfnamefont {M.}~\bibnamefont {Backhouse}}, \bibinfo {author}
  {\bibfnamefont {C.~D.}\ \bibnamefont {Baird}}, \bibinfo {author}
  {\bibfnamefont {M.~D.}\ \bibnamefont {Balcazar}}, \bibinfo {author}
  {\bibfnamefont {N.}~\bibnamefont {Bourgeois}},\ and\ \bibinfo {author}
  {\bibnamefont {et~al.}},\ }\bibfield  {title} {\bibinfo {title} {Automation
  and control of laser wakefield accelerators using bayesian optimization},\
  }\bibfield  {journal} {\bibinfo  {journal} {Nat. Commun.}\ }\textbf {\bibinfo
  {volume} {11}},\ \href {https://doi.org/10.1038/s41467-020-20245-6}
  {10.1038/s41467-020-20245-6} (\bibinfo {year} {2020})\BibitemShut {NoStop}%
\bibitem [{\citenamefont {Engel}(2003)}]{Engel2003}%
  \BibitemOpen
  \bibfield  {author} {\bibinfo {author} {\bibfnamefont {S.}~\bibnamefont
  {Engel}},\ }\bibfield  {title} {\bibinfo {title} {{Commissioning and
  operation of DRAGON}},\ }\href
  {https://doi.org/10.1016/S0168-583X(02)01910-9} {\bibfield  {journal}
  {\bibinfo  {journal} {Nucl. Instrum. Methods Phys. Res. B}\ }\textbf
  {\bibinfo {volume} {204}},\ \bibinfo {pages} {154} (\bibinfo {year}
  {2003})}\BibitemShut {NoStop}%
\bibitem [{\citenamefont {Couder}\ \emph {et~al.}(2008)\citenamefont {Couder},
  \citenamefont {Berg}, \citenamefont {G{\"{o}}rres}, \citenamefont {LeBlanc},
  \citenamefont {Lamm}, \citenamefont {Stech}, \citenamefont {Wiescher},\ and\
  \citenamefont {Hinnefeld}}]{Couder2008}%
  \BibitemOpen
  \bibfield  {author} {\bibinfo {author} {\bibfnamefont {M.}~\bibnamefont
  {Couder}}, \bibinfo {author} {\bibfnamefont {G.~P.~A.}\ \bibnamefont {Berg}},
  \bibinfo {author} {\bibfnamefont {J.}~\bibnamefont {G{\"{o}}rres}}, \bibinfo
  {author} {\bibfnamefont {P.~J.}\ \bibnamefont {LeBlanc}}, \bibinfo {author}
  {\bibfnamefont {L.~O.}\ \bibnamefont {Lamm}}, \bibinfo {author}
  {\bibfnamefont {E.}~\bibnamefont {Stech}}, \bibinfo {author} {\bibfnamefont
  {M.}~\bibnamefont {Wiescher}},\ and\ \bibinfo {author} {\bibfnamefont
  {J.}~\bibnamefont {Hinnefeld}},\ }\bibfield  {title} {\bibinfo {title}
  {{Design of the recoil mass separator St. GEORGE}},\ }\href
  {https://doi.org/10.1016/j.nima.2007.11.069} {\bibfield  {journal} {\bibinfo
  {journal} {Nucl. Instrum. Methods Phys. Res. A}\ }\textbf {\bibinfo {volume}
  {587}},\ \bibinfo {pages} {35} (\bibinfo {year} {2008})}\BibitemShut
  {NoStop}%
\bibitem [{\citenamefont {Meisel}\ \emph {et~al.}(2017)\citenamefont {Meisel},
  \citenamefont {Moran}, \citenamefont {Gilardy}, \citenamefont {Schmitt},
  \citenamefont {Seymour},\ and\ \citenamefont {Couder}}]{Meisel2017}%
  \BibitemOpen
  \bibfield  {author} {\bibinfo {author} {\bibfnamefont {Z.}~\bibnamefont
  {Meisel}}, \bibinfo {author} {\bibfnamefont {M.~T.}\ \bibnamefont {Moran}},
  \bibinfo {author} {\bibfnamefont {G.}~\bibnamefont {Gilardy}}, \bibinfo
  {author} {\bibfnamefont {J.}~\bibnamefont {Schmitt}}, \bibinfo {author}
  {\bibfnamefont {C.}~\bibnamefont {Seymour}},\ and\ \bibinfo {author}
  {\bibfnamefont {M.}~\bibnamefont {Couder}},\ }\bibfield  {title} {\bibinfo
  {title} {{Energy acceptance of the St. GEORGE recoil separator}},\ }\href
  {https://doi.org/10.1016/j.nima.2017.01.035} {\bibfield  {journal} {\bibinfo
  {journal} {Nucl. Instrum. Methods Phys. Res. A}\ }\textbf {\bibinfo {volume}
  {850}},\ \bibinfo {pages} {48} (\bibinfo {year} {2017})}\BibitemShut
  {NoStop}%
\bibitem [{\citenamefont {Kushner}(1964)}]{Kushner1964ANM}%
  \BibitemOpen
  \bibfield  {author} {\bibinfo {author} {\bibfnamefont {H.~J.}\ \bibnamefont
  {Kushner}},\ }\bibfield  {title} {\bibinfo {title} {A new method of locating
  the maximum point of an arbitrary multipeak curve in the presence of noise},\
  }\href@noop {} {\bibfield  {journal} {\bibinfo  {journal} {J. Basic Eng.}\
  }\textbf {\bibinfo {volume} {86}},\ \bibinfo {pages} {97} (\bibinfo {year}
  {1964})}\BibitemShut {NoStop}%
\bibitem [{\citenamefont {Mo{\v{c}}kus}(1975)}]{Mockus1975}%
  \BibitemOpen
  \bibfield  {author} {\bibinfo {author} {\bibfnamefont {J.}~\bibnamefont
  {Mo{\v{c}}kus}},\ }\bibfield  {title} {\bibinfo {title} {On bayesian methods
  for seeking the extremum},\ }in\ \href@noop {} {\emph {\bibinfo {booktitle}
  {Optimization Techniques IFIP Technical Conference Novosibirsk, July 1--7,
  1974}}},\ \bibinfo {editor} {edited by\ \bibinfo {editor} {\bibfnamefont
  {G.~I.}\ \bibnamefont {Marchuk}}}\ (\bibinfo  {publisher} {Springer Berlin
  Heidelberg},\ \bibinfo {address} {Berlin, Heidelberg},\ \bibinfo {year}
  {1975})\ pp.\ \bibinfo {pages} {400--404}\BibitemShut {NoStop}%
\bibitem [{\citenamefont {Brochu}\ \emph {et~al.}(2010)\citenamefont {Brochu},
  \citenamefont {Cora},\ and\ \citenamefont {de~Freitas}}]{brochu2010tutorial}%
  \BibitemOpen
  \bibfield  {author} {\bibinfo {author} {\bibfnamefont {E.}~\bibnamefont
  {Brochu}}, \bibinfo {author} {\bibfnamefont {V.~M.}\ \bibnamefont {Cora}},\
  and\ \bibinfo {author} {\bibfnamefont {N.}~\bibnamefont {de~Freitas}},\
  }\href@noop {} {\bibinfo {title} {A tutorial on bayesian optimization of
  expensive cost functions, with application to active user modeling and
  hierarchical reinforcement learning}} (\bibinfo {year} {2010}),\ \Eprint
  {https://arxiv.org/abs/1012.2599} {arXiv:1012.2599} \BibitemShut {NoStop}%
\bibitem [{\citenamefont {Rasmussen}\ and\ \citenamefont
  {Williams}(2018)}]{Rasmussen2018}%
  \BibitemOpen
  \bibfield  {author} {\bibinfo {author} {\bibfnamefont {C.~E.}\ \bibnamefont
  {Rasmussen}}\ and\ \bibinfo {author} {\bibfnamefont {C.~K.~I.}\ \bibnamefont
  {Williams}},\ }\href {https://doi.org/10.7551/mitpress/3206.001.0001} {\emph
  {\bibinfo {title} {Gaussian Processes for Machine Learning}}}\ (\bibinfo
  {publisher} {MIT Press, Cambridge},\ \bibinfo {year} {2018})\BibitemShut
  {NoStop}%
\bibitem [{\citenamefont {Cox}\ and\ \citenamefont {John}(1997)}]{Cox97}%
  \BibitemOpen
  \bibfield  {author} {\bibinfo {author} {\bibfnamefont {D.~D.}\ \bibnamefont
  {Cox}}\ and\ \bibinfo {author} {\bibfnamefont {S.}~\bibnamefont {John}},\
  }\bibfield  {title} {\bibinfo {title} {{SDO}: A statistical method for global
  optimization},\ }in\ \href@noop {} {\emph {\bibinfo {booktitle} {in
  Multidisciplinary Design Optimization: State-of-the-Art}}}\ (\bibinfo {year}
  {1997})\ pp.\ \bibinfo {pages} {315--329}\BibitemShut {NoStop}%
\bibitem [{\citenamefont {{GPy}}(2012)}]{GPy}%
  \BibitemOpen
  \bibfield  {author} {\bibinfo {author} {\bibnamefont {{GPy}}},\ }\href@noop
  {} {\bibinfo {title} {{GPy}: A gaussian process framework in python}},\
  \bibinfo {howpublished} {\url{http://github.com/SheffieldML/GPy}} (\bibinfo
  {year} {since 2012})\BibitemShut {NoStop}%
\bibitem [{\citenamefont {{The GPyOpt authors}}(2016)}]{Authors2016}%
  \BibitemOpen
  \bibfield  {author} {\bibinfo {author} {\bibnamefont {{The GPyOpt
  authors}}},\ }\href@noop {} {\bibinfo {title} {{GPyOpt}: A bayesian
  optimization framework in python}},\ \bibinfo {howpublished}
  {\url{http://github.com/SheffieldML/GPyOpt}} (\bibinfo {year}
  {2016})\BibitemShut {NoStop}%
\bibitem [{\citenamefont {Newville}\ \emph {et~al.}(2017)\citenamefont
  {Newville}, \citenamefont {Lauer}, \citenamefont {Dchabot}, \citenamefont
  {Caswell}, \citenamefont {Péteut}, \citenamefont {Hartman}, \citenamefont
  {Rokvintar}, \citenamefont {Clarken}, \citenamefont {Allan},\ and\
  \citenamefont {Birke}}]{Newville_2017}%
  \BibitemOpen
  \bibfield  {author} {\bibinfo {author} {\bibfnamefont {M.}~\bibnamefont
  {Newville}}, \bibinfo {author} {\bibfnamefont {K.}~\bibnamefont {Lauer}},
  \bibinfo {author} {\bibnamefont {Dchabot}}, \bibinfo {author} {\bibfnamefont
  {T.~A.}\ \bibnamefont {Caswell}}, \bibinfo {author} {\bibfnamefont
  {A.}~\bibnamefont {Péteut}}, \bibinfo {author} {\bibfnamefont
  {S.}~\bibnamefont {Hartman}}, \bibinfo {author} {\bibnamefont {Rokvintar}},
  \bibinfo {author} {\bibfnamefont {R.}~\bibnamefont {Clarken}}, \bibinfo
  {author} {\bibfnamefont {D.}~\bibnamefont {Allan}},\ and\ \bibinfo {author}
  {\bibfnamefont {T.}~\bibnamefont {Birke}},\ }\href
  {https://doi.org/10.5281/zenodo.883159} {\bibinfo {title} {pyepics/pyepics
  3.2.7}} (\bibinfo {year} {2017})\BibitemShut {NoStop}%
\bibitem [{\citenamefont {Hanuka}\ \emph {et~al.}(2021)\citenamefont {Hanuka},
  \citenamefont {Huang}, \citenamefont {Shtalenkova}, \citenamefont {Kennedy},
  \citenamefont {Edelen}, \citenamefont {Zhang}, \citenamefont {Lalchand},
  \citenamefont {Ratner},\ and\ \citenamefont {Duris}}]{hanuka2021}%
  \BibitemOpen
  \bibfield  {author} {\bibinfo {author} {\bibfnamefont {A.}~\bibnamefont
  {Hanuka}}, \bibinfo {author} {\bibfnamefont {X.}~\bibnamefont {Huang}},
  \bibinfo {author} {\bibfnamefont {J.}~\bibnamefont {Shtalenkova}}, \bibinfo
  {author} {\bibfnamefont {D.}~\bibnamefont {Kennedy}}, \bibinfo {author}
  {\bibfnamefont {A.}~\bibnamefont {Edelen}}, \bibinfo {author} {\bibfnamefont
  {Z.}~\bibnamefont {Zhang}}, \bibinfo {author} {\bibfnamefont {V.~R.}\
  \bibnamefont {Lalchand}}, \bibinfo {author} {\bibfnamefont {D.}~\bibnamefont
  {Ratner}},\ and\ \bibinfo {author} {\bibfnamefont {J.}~\bibnamefont
  {Duris}},\ }\bibfield  {title} {\bibinfo {title} {Physics model-informed
  gaussian process for online optimization of particle accelerators},\ }\href
  {https://doi.org/10.1103/PhysRevAccelBeams.24.072802} {\bibfield  {journal}
  {\bibinfo  {journal} {Phys. Rev. Accel. Beams}\ }\textbf {\bibinfo {volume}
  {24}},\ \bibinfo {pages} {072802} (\bibinfo {year} {2021})}\BibitemShut
  {NoStop}%
\bibitem [{\citenamefont {Roussel}\ \emph {et~al.}(2021)\citenamefont
  {Roussel}, \citenamefont {Hanuka},\ and\ \citenamefont
  {Edelen}}]{roussel2020multiobjective}%
  \BibitemOpen
  \bibfield  {author} {\bibinfo {author} {\bibfnamefont {R.}~\bibnamefont
  {Roussel}}, \bibinfo {author} {\bibfnamefont {A.}~\bibnamefont {Hanuka}},\
  and\ \bibinfo {author} {\bibfnamefont {A.}~\bibnamefont {Edelen}},\
  }\bibfield  {title} {\bibinfo {title} {Multiobjective bayesian optimization
  for online accelerator tuning},\ }\href
  {https://doi.org/10.1103/PhysRevAccelBeams.24.062801} {\bibfield  {journal}
  {\bibinfo  {journal} {Phys. Rev. Accel. Beams}\ }\textbf {\bibinfo {volume}
  {24}},\ \bibinfo {pages} {062801} (\bibinfo {year} {2021})}\BibitemShut
  {NoStop}%
\end{thebibliography}

%

\end{document}